# Thalamic nuclei segmentation from T$_1$-weighted MRI: unifying and benchmarking state-of-the-art methods with young and old cohorts

Brendan Williams[1,2†], Dan Nguyen[3], Julie P. Vidal[4,5], Manojkumar Saranathan[3†], for the Alzheimer's Disease Neuroimaging Initiative*

1. Centre for Integrative Neuroscience and Neurodynamics, University of Reading, United Kingdom
2. School of Psychology and Clinical Language Sciences, University of Reading, United Kingdom
3. Department of Radiology, University of Massachusetts Chan Medical School, Worcester, MA, United States
4. CNRS, CerCo (Centre de Recherche Cerveau et Cognition) - Université Paul Sabatier, Toulouse, France,
5. INSERM, ToNiC (Toulouse NeuroImaging Center) - Université Paul Sabatier, Toulouse, France

\* Data used in preparation of this article were obtained from the Alzheimer's Disease Neuroimaging Initiative (ADNI) database (adni.loni.usc.edu). As such, the investigators within the ADNI contributed to the design and implementation of ADNI and/or provided data but did not participate in analysis or writing of this report. A complete listing of ADNI investigators can be found at: http://adni.loni.usc.edu/wp-content/uploads/how_to_apply/ADNI_Acknowledgement_List.pdf

† Corresponding authors:

Brendan Williams, Centre for Integrative Neuroscience and Neurodynamics, Harry Pitt Building, University of Reading, Reading, Berkshire, United Kingdom. b.williams3@reading.ac.uk.

Manojkumar Saranathan, University of Massachusetts Medical School, 55 Lake Ave North, Worcester, Massachusetts, United States of America. manojkumar.saranathan@umassmed.edu.

# Abstract

The thalamus and its constituent nuclei are critical for a broad range of cognitive, linguistic, and sensorimotor processes, and are implicated in many neurological and neurodegenerative conditions. However, the functional involvement and specificity of thalamic nuclei in human neuroimaging work is underappreciated and not well studied due, in part, to technical challenges of accurately identifying and segmenting nuclei. This challenge is further exacerbated by a lack of common nomenclature for comparing segmentation methods. Here, we use data from healthy young (Human Connectome Project, n=100) and older healthy adults, plus those with minor cognitive impairment and Alzheimer's disease (Alzheimer's Disease Neuroimaging Initiative, n=540), to benchmark four state of the art thalamic segmentation methods for $T_1$ MRI (FreeSurfer, HIPS-THOMAS, SCS-CNN, and $T_1$-THOMAS) under a single segmentation framework. Segmentations were compared using overlap and dissimilarity metrics to the Morel stereotaxic atlas, a widely accepted thalamic atlas. We also quantified each method's estimation of thalamic nuclear degeneration across Alzheimer's disease progression, and how accurately early and late mild cognitive impairment, and Alzheimer's disease could be distinguished from healthy controls. We show that the HIPS-THOMAS approach produced the most effective segmentations of individual thalamic nuclei relative to the Morel atlas, and was also most accurate in discriminating healthy controls from those with minor cognitive impairment and Alzheimer's disease using individual nucleus volumes. This latter result was different when using whole thalamus volumes, where SCS-CNN approach was most accurate in classifying healthy controls. This work is the first to systematically compare the efficacy of anatomical thalamic segmentation approaches under a unified nomenclature. We also provide recommendations of which segmentation method to use for studying the functional relevance of specific thalamic nuclei, based on their overlap and dissimilarity with the Morel atlas.

# Introduction

Accurate identification of thalamic nuclei across spatial scales is important due to their widespread involvement in an array of functions, including sensory perception, motor control, sleep and arousal, linguistic, memory, and cognitive processes [1]–[7]. Aberrant thalamic structure and function are also implicated in a broad range of neurological, neuropsychiatric, developmental, and neurodegenerative conditions, including epilepsy, schizophrenia, autism spectrum disorder, multiple sclerosis, and Alzheimer's disease [8]–[18]. However, most MRI-based analyses treat the thalamus as a homogenous entity, reducing sensitivity to thalamic nuclei specific effects. Furthermore, thalamic nuclei segmentation from anatomical $T_1$- and $T_2$-weighted MRI data has been hampered by suboptimal image contrast resulting in poor delineation of intrathalamic and whole thalamus boundaries. Instead, most thalamic nuclei segmentation methods, to date, have been based on Diffusion Tensor Imaging (DTI), which is limited by the lack of anisotropy in the largely grey-matter dominant thalamus, and functional MRI, which is limited by poor spatial resolution and distortion of the underlying echoplanar imaging acquisition [19]–[24]. As a result, these methods do not resolve small structures such as lateral and medial geniculate nuclei (LGN/MGN), and the anteroventral (AV) nucleus, which are critical for sensory and cognitive processing.

Due to its inclusion in most publicly available datasets and neuroimaging protocols, alongside its high isotropic spatial resolution (usually 1 mm or better), there has been a renewed interest in thalamic segmentation based on anatomical $T_1$-weighted ($T_1$w) MRI, despite its poor contrast in the thalamus. Recently introduced thalamic segmentation methods like the FreeSurfer Bayesian inference [25], and the THOMAS multi-atlas [26] approaches have been used to analyze data in disease states like Alzheimer's disease, alcohol use disorder, and multiple sclerosis [27]–[30]. While FreeSurfer primarily works on $T_1$ Magnetization Prepared RApid Gradient Echo (MPRAGE) data with the ability to incorporate secondary images with different image contrast [31], the original THOMAS algorithm [26] was optimized and validated using white-matter-nulled (WMn) MPRAGE, a special pulse sequence that nulls white-matter instead of cerebrospinal fluid (CSF) as in standard MPRAGE. THOMAS was recently adapted for conventional $T_1$w MRI in a modified method ($T_1$-THOMAS), using a mutual information (MI) metric for nonlinear registration and a majority voting (MV) algorithm for label fusion. However, $T_1$-THOMAS [32] was not as accurate compared to segmentations based on WMn-MPRAGE for several small nuclei, presumably due to loss of intrathalamic contrast and poor delineation of thalamic boundaries in $T_1$w MPRAGE contrast. To leverage the improved intrathalamic contrast of WMn imaging, a deep learning-based approach has been proposed using a first convolutional neural network (CNN) to synthesize WMn-MPRAGE-like images from $T_1$w-MRI and a second CNN to perform segmentation on the synthesized WMn images. This method, called synthesized contrast segmentation (SCS), was shown to be much more accurate than direct CNN segmentation of $T_1$w-MRI data [33]. Another recent method uses a robust histogram-based polynomial synthesis (HIPS) approach instead of a CNN for the synthesis of WMn-MPRAGE-like images, those synthesized images are then identically processed as in the original THOMAS method for thalamic nuclei segmentation [34]. This method also showed significant improvement in Dice and reduction in volume errors compared to $T_1$-THOMAS and was demonstrated to be more robust than the SCS-CNN when applied to data from higher field strengths or scanner manufacturers that were not part of the CNN training process [34].

A widely used reference guide for identifying thalamic nuclei is the Morel stereotaxic atlas [35], [36], which was developed using histological staining of five post-mortem brains from healthy older adults for the calcium binding proteins parvalbumin, calbindin D-28k, and calretinin to identify cyto- and myeloarchitectural features. Functional relevance was also considered during the development of the Morel atlas. More recently, the Morel atlas has been digitised (Krauth-Morel atlas) and made available in MNI-space for use as a potential reference in a wide variety of neuroimaging applications [37]. However, despite their claims of conformity with the Morel atlas, both FreeSurfer and THOMAS use different nomenclatures and definitions for thalamic nuclei and produce parcellations which differ qualitatively from each other. As a result, direct comparisons of these segmentation methods on the same datasets have not been reported. Here, we systematically compare four state-of-the-art methods for thalamic nuclei segmentation of $T_1$w MRI: FreeSurfer, HIPS-THOMAS, SCS-CNN, and $T_1$-THOMAS. We used data from healthy younger adults in the Human Connectome Project (HCP) to quantitatively compare segmentations from these methods against the Krauth-Morel atlas [37] in subject-space as well as MNI-space. We then analysed data from older adults from the Alzheimer's Disease Neuroimaging Initiative (ADNI) database to characterize thalamic atrophy as a function of disease status and assessed the accuracy of each of the methods in predicting Alzheimer's disease status using a receiver operating characteristic (ROC) analysis.

**Methods**

**Participants**

Anatomical $T_1$w-MRI data were sourced from two publicly available neuroimaging datasets: the Human Connectome Project (HCP) [38] and Alzheimer's Disease Neuroimaging Initiative (ADNI) (http://adni.loni.usc.edu). 100 subjects were pseudo-randomly selected from the HCP dataset as in previous work [39]. A subset of 540 subjects were selected from the ADNI dataset, comprising of participants who had undergone a Montreal Cognitive Assessment (MoCA) test and were scanned on 3T MRI using MPRAGE followed by successful image registration and segmentation (see [32] for further details). Subjects were classified as either healthy control (HC, 119 subjects), early minor cognitive impairment (EMCI, 208 subjects), late minor cognitive impairment (LMCI, 116 subjects) or Alzheimer's disease (AD, 97 subjects). EMCI and LMCI were classified based on subjective memory concern scores (either themselves, their partner, or a clinician) from the logical memory II subscale of the Wechsler Memory Scale – Revised (16+ years in education: EMCI scores 9-11, LMCI > 4 & ≤ 8; 8-15 years in education: EMCI scores 5-9, LMCI > 2 & ≤ 4; 0-7 years in education: EMCI scores 3-6, LMCI > 0 & ≤ 2), a Mini-Mental State Examination Score between 24-30, a Clinical Dementia Rating of 0.5 in the memory box, and sufficient cognitive and functional performance that would not make threshold for an AD diagnosis [32].

**Data processing**

Pre-processing:

HCP data (n=100) were pre-processed using the HCP minimal preprocessing pipelines [39], [40]. Firstly, $T_1w$ images were corrected for gradient distortions using a customised version of *gradient_nonlin_unwarp* in FreeSurfer, then each subject's two $T_1w$ scans were aligned using FSL FLIRT and averaged. The averaged $T_1w$ image was then registered to MNI-space using a 12 DOF affine registration with FLIRT, and a subset of 6 DOF transforms were used to align the anterior commissure, the anterior commissure – posterior commissure line, and the inter-hemispheric plane, while preserving the size and shape of the brain in native space. The skull was removed by inverting linear (FLIRT) and nonlinear (FNIRT) warps from anatomical to MNI-space, applying the warp to the MNI-space brain mask, and then applying the mask to the averaged $T_1w$ image. Finally, the image was corrected for readout distortion and biases in $B_1$ and $B_1^+$ fields. $T_1w$ MPRAGE datasets from ADNI (n=540) were directly processed using the different thalamic nuclei segmentation methods with no extra preprocessing steps. Note that the N4 bias correction to remove shading is incorporated inside the THOMAS and SCS-CNN pipelines.

Thalamic nuclei segmentation:

The four main thalamic segmentation schemes compared in this work are summarized in Figure 1 and described below.

**FreeSurfer**: HCP and ADNI data were segmented following methods described previously [25], [39]. Data processing was run using a Nipype pipeline integrating FSL (version 6.0.4) and FreeSurfer (version 7.1.1). Anatomical $T_1w$ images were first processed and parcellated using *recon-all* in FreeSurfer; the output of recon-all was used to initialise the parcellation of thalamic nuclei for anatomical data using the algorithm described by [25]. The parcellated thalamus was converted from FreeSurfer space to native anatomical space and changed from mgz to nii file format using *mri_label2vol* and *mri_convert* in FreeSurfer, respectively.

**SCS-CNN**: $T_1$w MRI datasets from both the HCP and ADNI databases were segmented using the dual CNN method described in [33] and run as a docker container obtained from Lavanya Umapathy. Briefly, the SCS-CNN method uses two CNNs- the first CNN was trained using patches from contemporaneously acquired $T_1$w and WMn-MPRAGE data is used to synthesize WMn-like images from $T_1$w data. The second CNN was trained using THOMAS labels from WMn-MPRAGE data and was used to segment the synthesised WMn-like image from the first CNN.

**THOMAS variants**: $T_1$-THOMAS is an adaptation of the original THOMAS method for $T_1$w-MRI that uses a mutual information metric for nonlinear registration and majority voting for label fusion [32]. HIPS-THOMAS incorporates a HIPS preprocessing step using information from the histograms and a plot of each voxel's intensities of $T_1$w and WMn-MPRAGE images to compute a polynomial approximation from a small subset of training images used to synthesize WMn-like images from $T_1$w [34]. More information and schematics can be found in the Supplementary methods and Supplemental Figure 1.

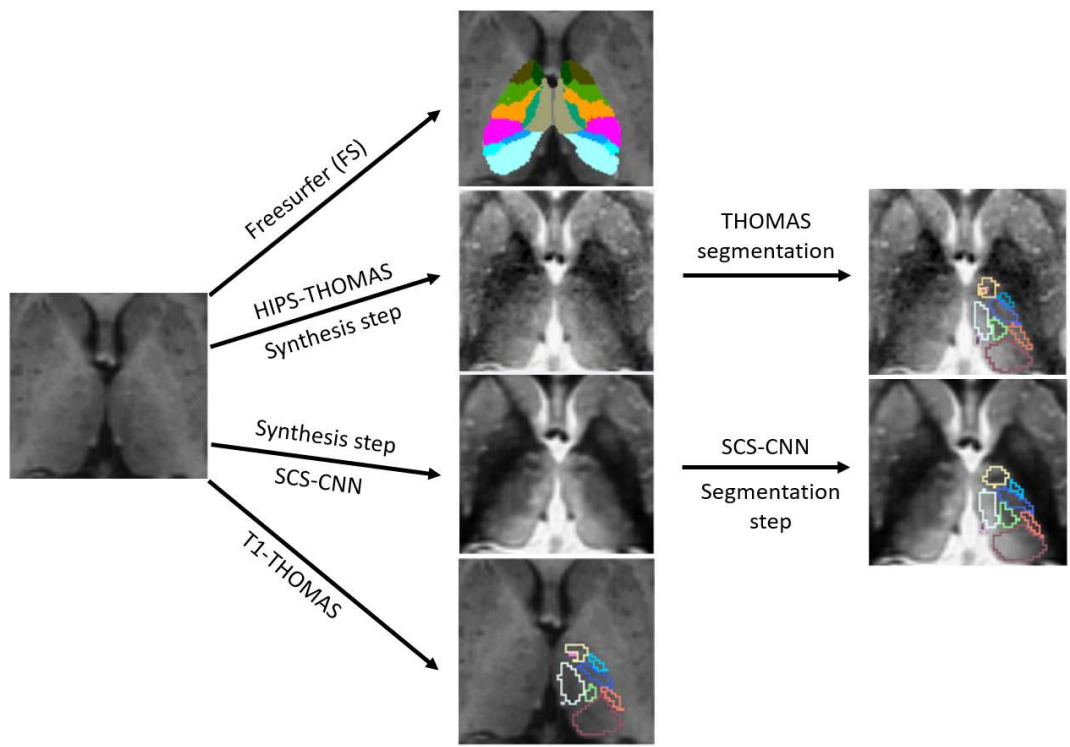

Figure 1: Overview of thalamic nuclei segmentation used for HCP and ADNI $T_1$w-MRI datasets showing the FreeSurfer, HIPS-THOMAS, SCS-CNN, and $T_1$-THOMAS schemes.

## Label synthesis and segmentation preparation

For FreeSurfer outputs and the Krauth-Morel atlas, thalamic nuclei were combined to match the Morel nomenclature used by THOMAS to generate 10 thalamic nuclei for subjects in both the HCP and ADNI datasets (Table 1). Note that habenula and mammillothalamic tract (MTT) were omitted as FreeSurfer did not segment those structures. Similarly, lateral nuclei such as lateral dorsal and intralaminar nuclei such as centrolateral which are not segmented by THOMAS were omitted. For the HCP subjects, rigid and affine transformations, and non-linear warps between native subject space and MNI-space were generated with the Advanced Normalization Tools (ANTs) package (Version 2.3.5, Ecphorella) [41]. These image transformations and warps were used to generate subject space versions of Krauth-Morel nuclei, and MNI-space versions of FreeSurfer, SCS-CNN, and HIPS-THOMAS segmented nuclei for comparison using nearest neighbour interpolation in ANTs. The MNI-space versions of the segmented nuclei were used to generate group-level probabilistic atlases [42].

Table 1 FreeSurfer and Krauth nuclei combined to match the Morel nomenclature used by THOMAS.

| THOMAS nuclei | FreeSurfer nuclei | Krauth-Morel nuclei |
|---|---|---|
| AV | AV | AV |
| VA | VAmc + VApc | VAmc + VApc |
| VLa | VLa | VLa |
| VLp | VLp | VLpd + VLpv + VLp |
| VPL | VPL | VPLa + VPLp |
| Pul | PuA + PuI + PuL+ PuM | PuA+ PuI+ PuL+ PuM |
| LGN | LGN | LGNmc + LGNpc |
| MGN | MGN | MGN |
| CM | CM | CM |
| MD-Pf | MDl+ MDm + Pf | Pf + sPf + MDmc + MDpc |

**Segmentation metrics**

Dice similarity coefficients (Dice) were used to compare the 10 segmented thalamic nuclei per hemisphere from each approach with the Krauth-Morel atlas in both subject space and in MNI-space for the HCP data. Dice is a widely used measure in image processing for assessing overlap, and is defined as:

$$Dice(S_x, S_y) = \frac{2|S_x \cap S_y|}{|S_x| + |S_y|}$$

where $|S_x \cap S_y|$ is the cardinality of the intersection between the segmentation and ground truth (this is equal to the number of true positives, or overlapping voxels), divided by the sum of the cardinality of the ground truth $|S_x|$ and the segmentation $|S_y|$ (equal to the sum of true positives, false positives, and false negatives) [39]. Using subject space data, we ranked each segmentation approach from best to worst (1 to 4) for each nuclei using paired sample t-test results. For the MNI-space analysis, a threshold of 0.25 was used to binarize the group-level probabilistic atlas. We used the following cut-offs to compare segmentations: Dice = 0 no agreement, 0 < Dice < 0.2 slight agreement, 0.2 ≤ Dice < 0.4 fair agreement, 0.4 ≤ Dice < 0.6 moderate agreement, 0.6 ≤ Dice < 0.8 substantial agreement, 0.8 ≤ Dice ≤ 1 almost perfect agreement [43]. We defined the best segmentation approach for each nucleus based on subject-space and MNI-space results. A single segmentation was defined as the best for a given nuclei if it had the highest MNI-space Dice coefficient and had a significantly higher Dice coefficient in the subject space analysis than other approaches. Two or more segmentations were defined jointly as being the best if they a) had non-significant differences at the subject level and different directions of Dice coefficients results for subject and MNI space data, or b) had significantly different scores at the subject level but the direction for the group Dice coefficient showed the inverse effect to the result of the null hypothesis significance test. If more than two segmentations met the previous criteria, then no segmentation approach was defined as the best overall.

We also calculated the Average Hausdorff Distance (AHD) to compare the 10 segmented thalamic nuclei with the Krauth-Morel atlas. The Average Hausdorff Distance is used as a measure of dissimilarity and can account for differences in isometry. Distance-based metrics are advantageous relative to overlap-metrics in situations where segmentations are small because overlap-based metrics disproportionately penalise errors in smaller than larger segmentations [44] as is the case with thalamic nuclei. In relation to image segmentation, the Hausdorff Distance can be defined as the minimum number of voxels between a point in segmentation X and a point in segmentation Y. Therefore, the Average Hausdorff Distance is the average minimum distance between all points in segmentation A and segmentation B in voxels. The Average Hausdorff Distance is defined as:

$$\text{Average Hausdorff Distance}(A, B) = \max(d(A,B), d(B,A))$$

$$d(A, B) = \frac{1}{N} \sum_{a \in A} \min_{b \in B} \|a - b\|$$

where $d(A, B)$ is the average minimum distance ($min\|a - b\|$) from voxels in the ground truth ($A$) to the segmentation ($B$), $d(B, A)$ is the average minimum distance ($min\|a - b\|$) from voxels in the segmentation ($B$) to the ground truth ($A$). The Average Hausdorff Distance is then the maximum of either of these two average distance measures in voxels [39].

**Statistical analysis**

We performed statistical analyses using R and jamovi. We used two-way ANOVAs to compare segmentation metrics (Dice coefficient and Average Hausdorff Distance) between different segmentation methods in subject space, testing for main effects of segmentation methods (FreeSurfer, HIPS-THOMAS, SCS-CNN, and $T_1$-THOMAS), hemisphere (left and right), and for interactions between segmentation methods and hemispheres. Posthoc t-tests (Bonferroni corrected) were used to compare between segmentation methods for each nucleus. We used ANCOVAs to compare nuclei volumes for each segmentation method using the ADNI dataset to test for main effects of group (HC, EMCI, LMCI, AD), and included age, biological sex, years of education, and intracranial volume (eTIV output of FreeSurfer) as covariates. Dunnett's test were used for posthoc analyses to compare HC with EMCI, LMCI, and AD groups. We obtained least squares estimates for volumes after adjusting for covariates, and effects with adjusted p < 0.05 were considered statistically significant. To calculate effect sizes for each pair-wise comparison we computed Cohen's d. Lastly, we performed a Receiver Operating Characteristic (ROC) analysis for the ADNI dataset using logistic regression to quantify the ability of each segmentation method (nuclei volumes) to discriminate EMCI, LMCI, and AD from HC. We calculated area under the curve (AUC) values for each of the three scenarios for the different segmentation methods. AUCs were also computed using whole thalamus volumes alone for comparison.

## Results

### Human Connectome Project

**Subject-space analysis** Two-way ANOVAs found significant main effects of segmentation approach and hemisphere for all nuclei (except for VPL, which had a non-significant main effect of hemisphere), and a significant interaction between segmentation approach and hemisphere for all nuclei (Supplementary Table 1). Posthoc t-tests (Bonferroni corrected) for main effects of segmentation approach on Dice coefficients are summarized in Figure 2 for left thalamic nuclei and Figure 3 for right thalamic nuclei. To compare segmentation approaches based on the subject-space Dice coefficients we ranked each segmentation approach from best (1) to worst (4) based on the posthoc t-tests for main effects of segmentation approach (Table 2). Overall, HIPS-THOMAS had the best (lowest) mean ranking (L 1.9, R 1.8), followed by the FreeSurfer (L 2.1, R 2.5) and SCS-CNN approaches (L 2.2, R 2.5), and then THOMAS (L 3.3, R 3). HIPS-THOMAS also had the lowest overall variation in ranking, as shown by the smaller standard deviation compared to the other approaches. We calculated the Average Hausdorff Distance in voxels between the segmented nuclei and the corresponding nuclei from the Krauth-Morel atlas in subject space for left (Figure 4) and right (Figure 5) nuclei. We also used two-way ANOVAs to compare Average Hausdorff Distances between segmentation approaches and hemisphere for each thalamic nucleus. We found a significant main effect of segmentation approach for VA ($F(1, 98.15)=9.886$, $p=0.002$, $\eta_g^2=0.037$), VLp ($F(1, 98.26)=4.052$, $p=0.047$, $\eta_g^2=0.015$), LGN ($F(1.01, 98.89)=5.02$, $p=0.027$,

$\eta_g^2$=0.019), MGN ($F$(1.02, 99.95)=10.816, $p$=0.001, $\eta_g^2$=0.04), CM ($F$(1, 98.32)=6.062, $p$=0.015, $\eta_g^2$=0.023), and MD-Pf ($F$(1, 98.17)=4.28, $p$=0.041, $\eta_g^2$=0.016), but no significant main effects of segmentation approach for AV, VLa, VPL, or Pul. We found a significant main effect of hemisphere for LGN ($F$(1, 98)=5.674, $p$=0.019, $\eta_g^2$=0.007) and MGN ($F$(1, 98)=11.223, $p$=0.001, $\eta_g^2$=0.014), but not for other thalamic nuclei (supplementary figures 2 & 3). We found no significant interactions between segmentation approach and hemisphere. Posthoc t-tests (Bonferroni corrected) for main effects of segmentation approach on Average Hausdorff Distances are summarized in supplementary Figures 2 and 3 for left and right thalamic nuclei, respectively. Like for Dice, we ranked each segmentation approach from best (lowest AHD) to worst (highest AHD) (supplementary table 2). Using only nuclei where there was a main effect of segmentation approach on AHD, HIPS-THOMAS had the best (lowest) mean ranking (L & R 1.67), followed by FreeSurfer in the left hemisphere (1.83) and SCS-CNN in the right hemisphere (2). Both FreeSurfer and SCS-CNN had rankings of 2.3 in the right and left hemispheres, respectively. Bilaterally, THOMAS had the worst mean ranking (L 3.83, R 3.3). The lowest variation in rankings was found for THOMAS in the left hemisphere (SD = 0.37); the next lowest variation (SD = 0.75) was found bilaterally for HIPS-THOMAS, and in the left and right hemispheres for SCS-CNN and THOMAS, respectively. Bilaterally, FreeSurfer had the greatest variation in rankings (L SD = 1.21, R SD = 0.94).

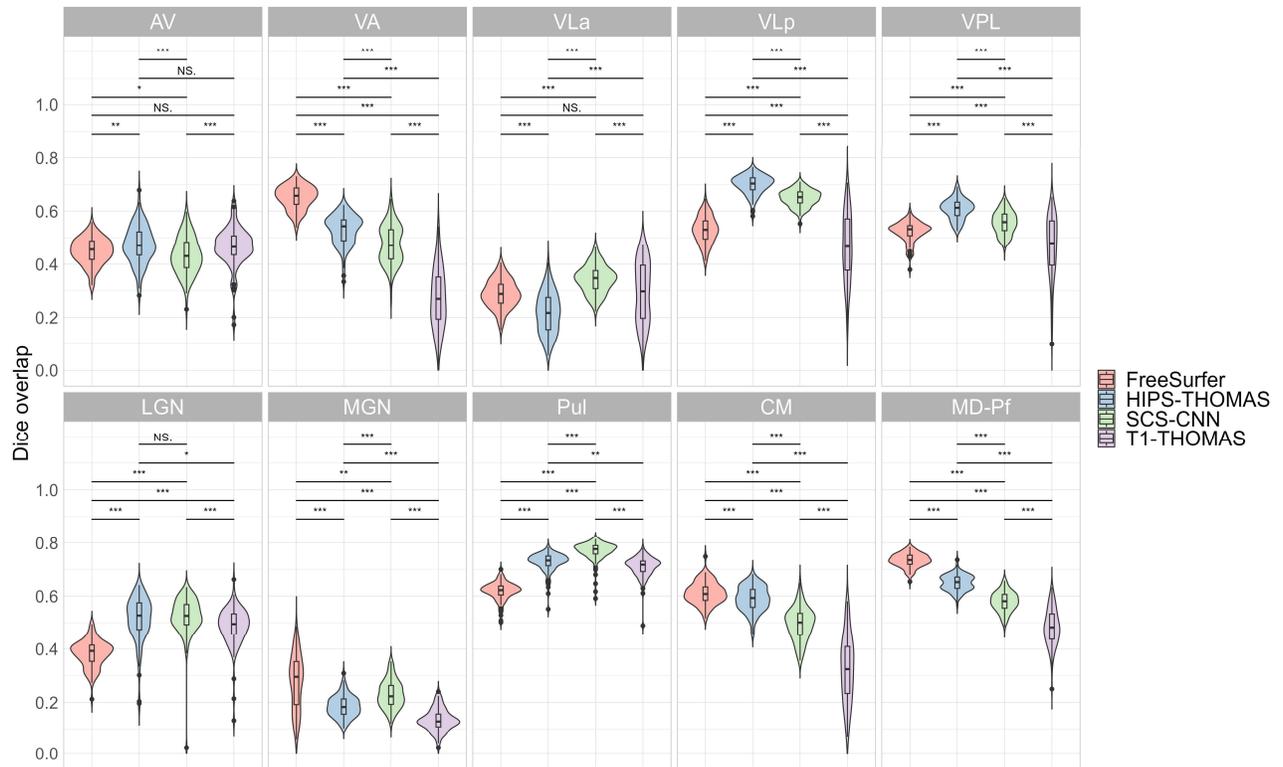

Figure 2: Violin plots of Dice overlap between left hemisphere nuclei segmented from Human Connectome Project data using FreeSurfer, HIPS-THOMAS, CNN-SCS, and $T_1$-THOMAS approaches. Posthoc t-test results (Bonferroni corrected) are presented to show pairwise difference between segmentation approaches for each nucleus (*p<0.05, **p<0.01, ***p<0.001).

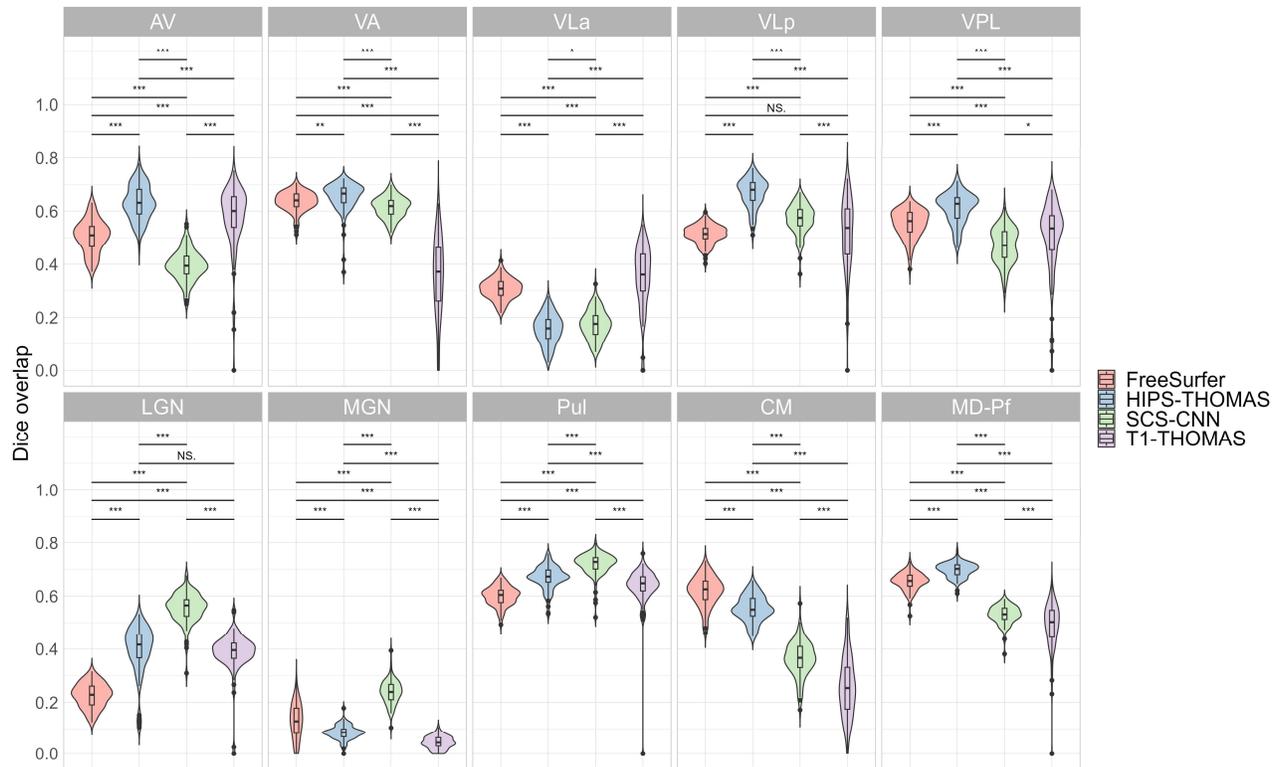

Figure 3 Violin plots of Dice overlap between right hemisphere nuclei segmented from Human Connectome Project data using FreeSurfer, HIPS-THOMAS, CNN-SCS, and $T_1$-THOMAS approaches. Posthoc t-test results (Bonferroni corrected) are presented to show pairwise difference between segmentation approaches for each nucleus (*p<0.05, **p<0.01, ***p<0.001).

Table 2: Rankings for each of the segmentation approaches for each nucleus, based on their mean Dice values and significance levels (see Figure 2 and Figure 3 for further details). Segmentations were ranked from best (1) to worst (4)

| | Left | | | | Right | | | |
|---|---|---|---|---|---|---|---|---|
| **Nucleus** | FreeSurfer | HIPS-THOMAS | SCS-CNN | T1-THOMAS | FreeSurfer | HIPS-THOMAS | SCS-CNN | T1-THOMAS |
| **AV** | 1 | 1 | 4 | 1 | 3 | 1 | 4 | 2 |
| **VA** | 1 | 2 | 3 | 4 | 2 | 1 | 3 | 4 |
| **Vla** | 2 | 4 | 1 | 2 | 2 | 4 | 3 | 1 |
| **VLp** | 3 | 1 | 2 | 4 | 3 | 1 | 2 | 3 |
| **VPL** | 3 | 1 | 2 | 4 | 2 | 1 | 4 | 3 |
| **LGN** | 4 | 1 | 1 | 3 | 4 | 2 | 1 | 2 |
| **MGN** | 1 | 3 | 2 | 4 | 2 | 3 | 1 | 4 |
| **Pul** | 4 | 2 | 1 | 3 | 4 | 2 | 1 | 3 |
| **CM** | 1 | 2 | 3 | 4 | 1 | 2 | 3 | 4 |
| **MD-Pf** | 1 | 2 | 3 | 4 | 2 | 1 | 3 | 4 |
| **Mean** | 2.1 | 1.9 | 2.2 | 3.3 | 2.5 | 1.8 | 2.5 | 3 |
| **SD** | 1.164 | 0.899 | 0.934 | 0.958 | 0.879 | 0.934 | 1.066 | 0.953 |

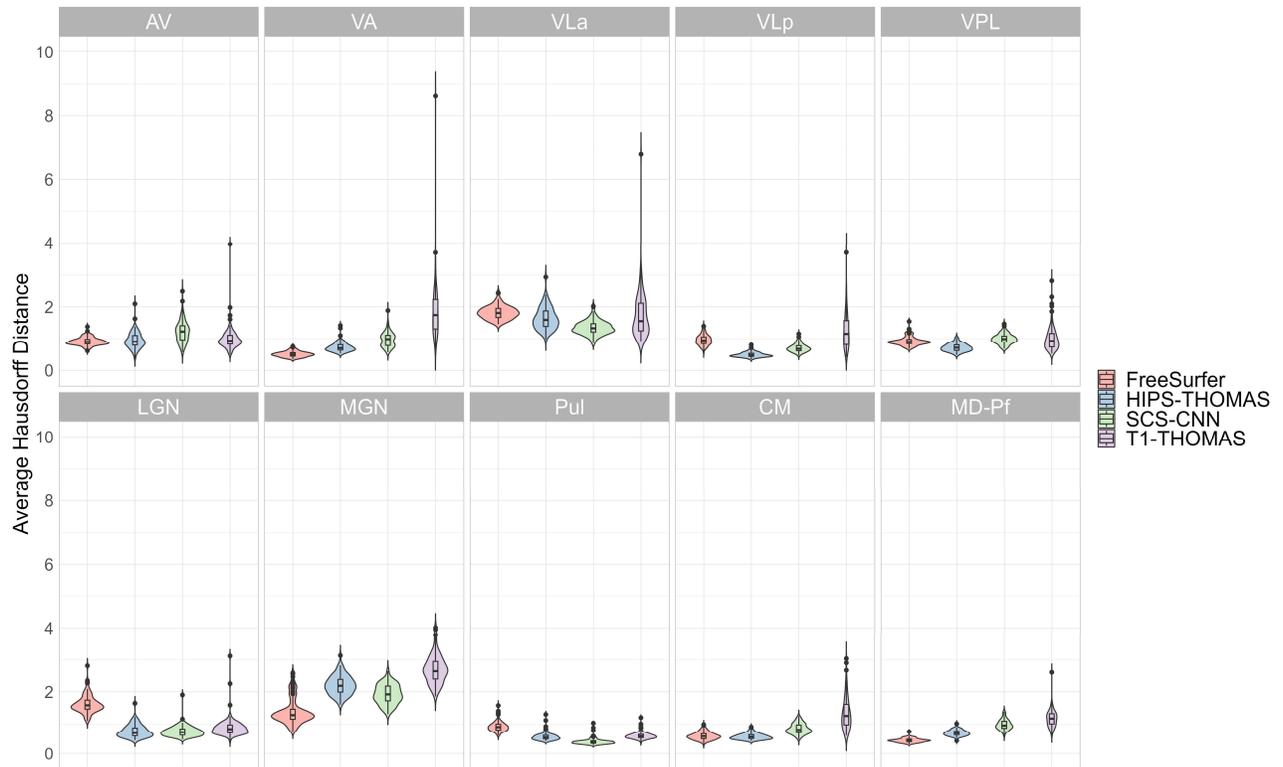

Figure 4: Violin plots of Average Hausdorff Distance for left hemisphere nuclei segmented from Human Connectome Project data using FreeSurfer, HIPS-THOMAS, CNN-SCS, and $T_1$-THOMAS approaches. Significant main effects of segmentation approach were found for VA, VLp, LGN, MGN, CM, and Md-Pf. Posthoc t-test results (Bonferroni corrected) are presented in supplementary Figure 2.

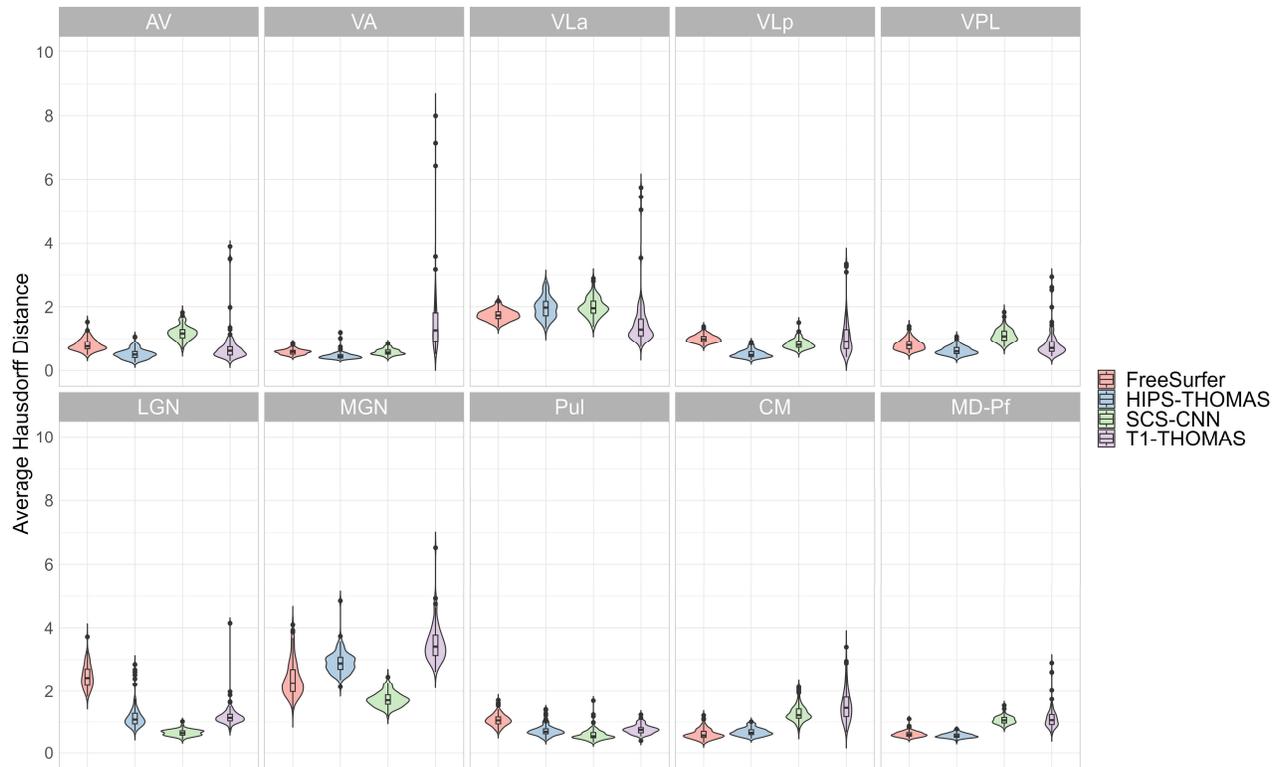

Figure 5: Violin plots of Average Hausdorff Distance for right hemisphere nuclei segmented from Human Connectome Project data using FreeSurfer, HIPS-THOMAS, CNN-SCS, and $T_1$-THOMAS approaches. Significant main effects of segmentation approach were found for VA, VLp, LGN, MGN, CM, and Md-Pf. Posthoc t-test results (Bonferroni corrected) are presented in supplementary Figure 2.

**MNI-space analysis**: Dice and Hausdorff-distance metrics computed in MNI-space are tabulated for the left and right hemispheres for each of the four methods are summarised in Figure 6 and Figure 7, respectively. An overview of the classification of Dice coefficients is presented in Table 2. Although SCS-CNN was the only method to have a nucleus with the "almost perfect agreement" classification, it had a wider distribution of classifications. Similar distributions were seen for FreeSurfer and $T_1$-THOMAS, while HIPS-THOMAS was the only approach to have eight of ten nuclei in the "substantial agreement" classification group. For the left hemisphere, HIPS-THOMAS had the maximum Dice for 4 nuclei (VLp, VPL, LGN, and CM), FreeSurfer for 4 nuclei (AV, VA, MGN, and MD-Pf) and SCS-CNN for 2 nuclei (VLa, Pul). For the right hemisphere, HIPS-THOMAS had the maximum Dice for 5 nuclei (AV, VA, VLp, VPL, and MD), FreeSurfer for 2 nuclei (VLa, CM) and SCS-CNN for 2 nuclei (MGN, Pul). Nuclei had smaller Average Hausdorff Distances along the diagonal (same nuclei) across segmentation approaches, except for left VLa for FreeSurfer, which bilaterally was closer to VLp, suggesting FreeSurfer is not able to distinguish well between VLa and VLp.

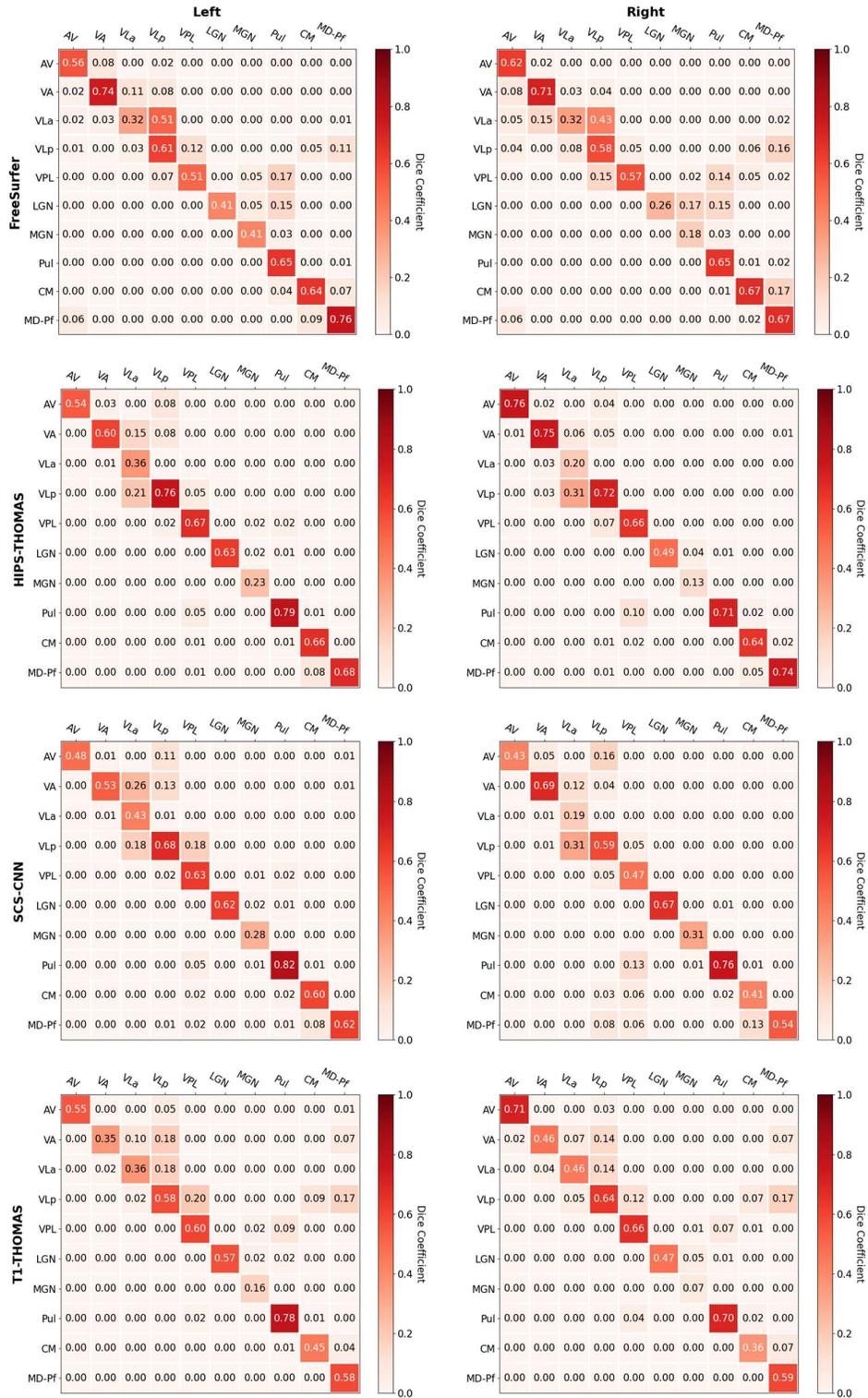

Figure 6: Group-level Dice overlap coefficients for nuclei segmented from Human Connectome Project data using FreeSurfer, HIPS-THOMAS, CNN-SCS, and $T_1$-THOMAS approaches in MNI-space.

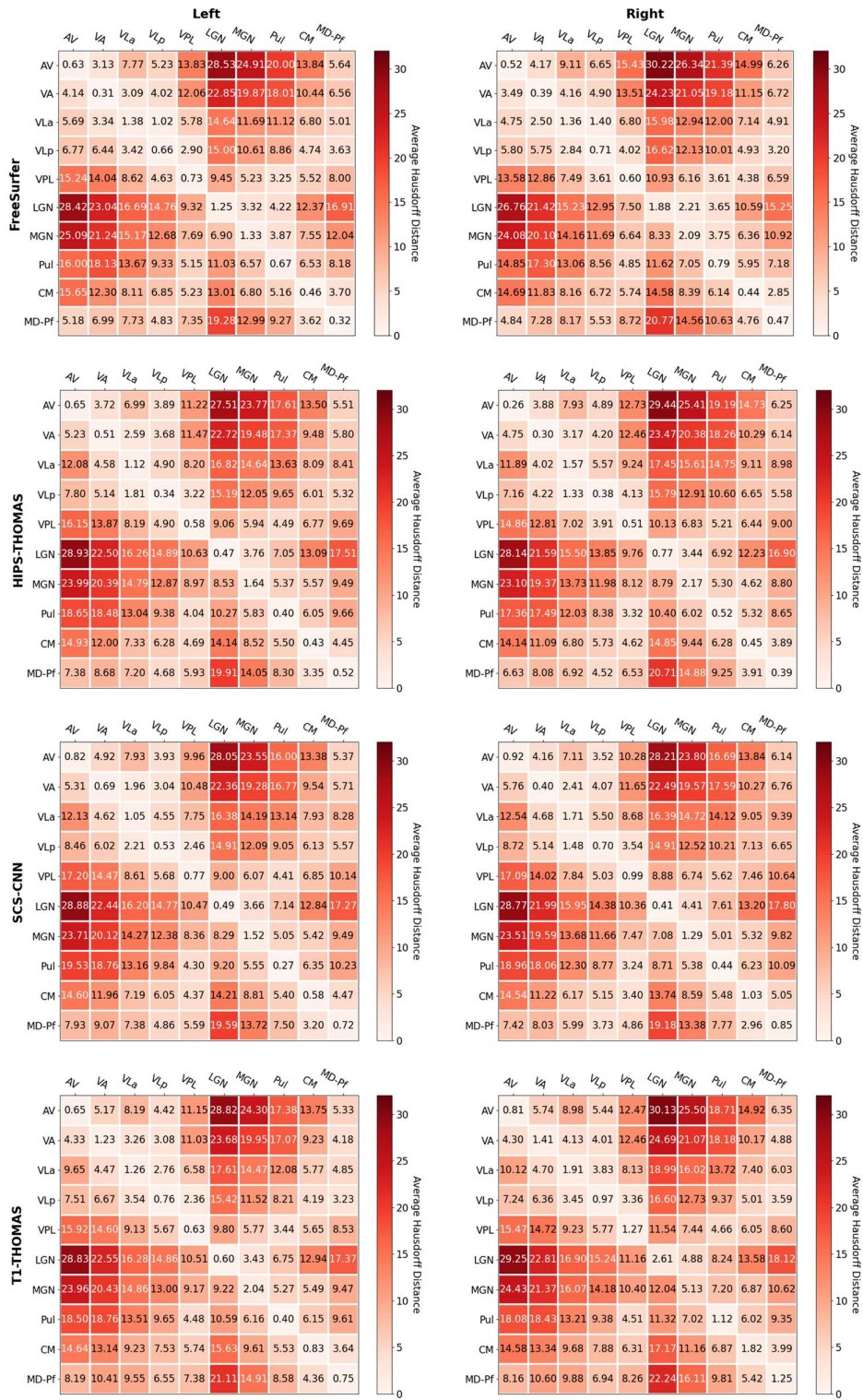

Figure 7 Group-level Average Hausdorff Distance for nuclei segmented from Human Connectome Project data using FreeSurfer, HIPS-THOMAS, CNN-SCS, and T1-THOMAS approaches in MNI-space.

Table 3 Classification of Dice coefficients for each nucleus across segmentation approaches. Nuclei where a single method produced the highest MNI space Dice result for both hemispheres are presented in bold; nuclei where different methods produced the highest MNI space Dice result for each hemisphere are presented in italics.

|  | Laterality | FreeSurfer | HIPS-THOMAS | SCS-CNN | T1-THOMAS |
|---|---|---|---|---|---|
| Almost perfect agreement | Unilateral |  |  | **Pul** |  |
| Substantial agreement | Bilateral | VA, Pul, *CM, MD-Pf* | **VLp, VPL**, Pul, *CM, MD-Pf* | **LGN** | Pul |
|  | Unilateral | *AV*, VLp | *AV, VA*, LGN | VA, VLp VPL, MD-Pf | AV, VLp, VPL |
| Moderate agreement | Bilateral | VPL |  | AV, CM | LGN, MD-Pf |
|  | Unilateral | LGN, *MGN* |  | *VLa* | VA, VLa, CM |
| Fair agreement | Bilateral | *VLa* |  | MGN |  |
|  | Unilateral |  | VLa, MGN |  |  |
| Slight agreement | Bilateral |  |  |  | MGN |

Interim summary:

The "best" method for each nucleus based on subject and MNI-space Dice and AHD coefficients are graphically depicted in Figure 8. The best performing methods were broadly similar for both subject and MNI-space coefficients, as reflected by the selection of a single nucleus in the right hemisphere, and all but three nuclei in the left hemisphere. In the left hemisphere HIPS-THOMAS and FreeSurfer/SCS-CNN were joint best for CM and LGN respectively, while several nuclei showed comparable Dice coefficients for AV.

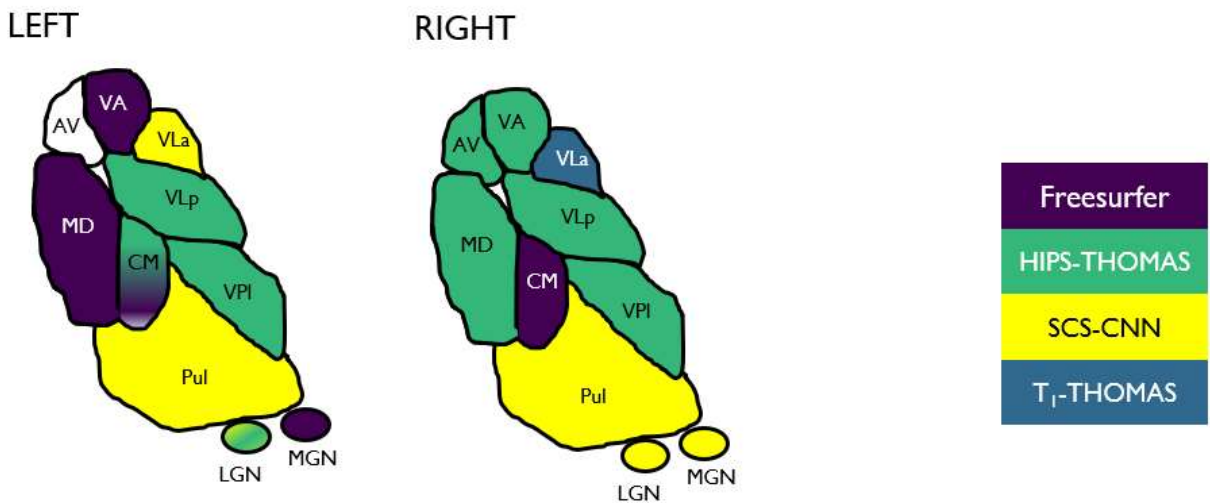

Figure 8 Best segmentation approach for each thalamic nuclei in each hemisphere, based on subject space and group Dice coefficients.

**Alzheimer's Disease Neuroimaging Initiative**

Figure 9 and Figure 10 show thalamic nuclei atrophy colorized using Cohen's d for the left and right hemispheres respectively, for HC-EMCI, HC-LMCI, and HC-AD comparisons. Only nuclei with statistically significant differences in the ANCOVA tests are coloured. The Cohen's d provides a dimensionless metric for comparisons across methods. The progression of atrophy from EMCI to LMCI is captured nicely by SCS-CNN and HIPS-THOMAS while FreeSurfer and $T_1$-THOMAS do not exhibit a clear progression from EMCI to LMCI. Note that while SCS-CNN displays the progression it was with reduced effect sizes.

The results of the ROC analyses are summarized in Table 4. AUC values for discrimination of AD and HC for FreeSurfer, HIPS-THOMAS, SCS-CNN, and $T_1$-THOMAS using all the individual thalamic nuclei volumes (adjusted for ICV/age) and whole thalamus volumes (for comparison) are shown. Classification accuracy using thalamic nuclei volumes was most accurate using HIPS-THOMAS while SCS-CNN was the most accurate when classifying using whole thalamus volumes, albeit with smaller AUCs for all three disease stages.

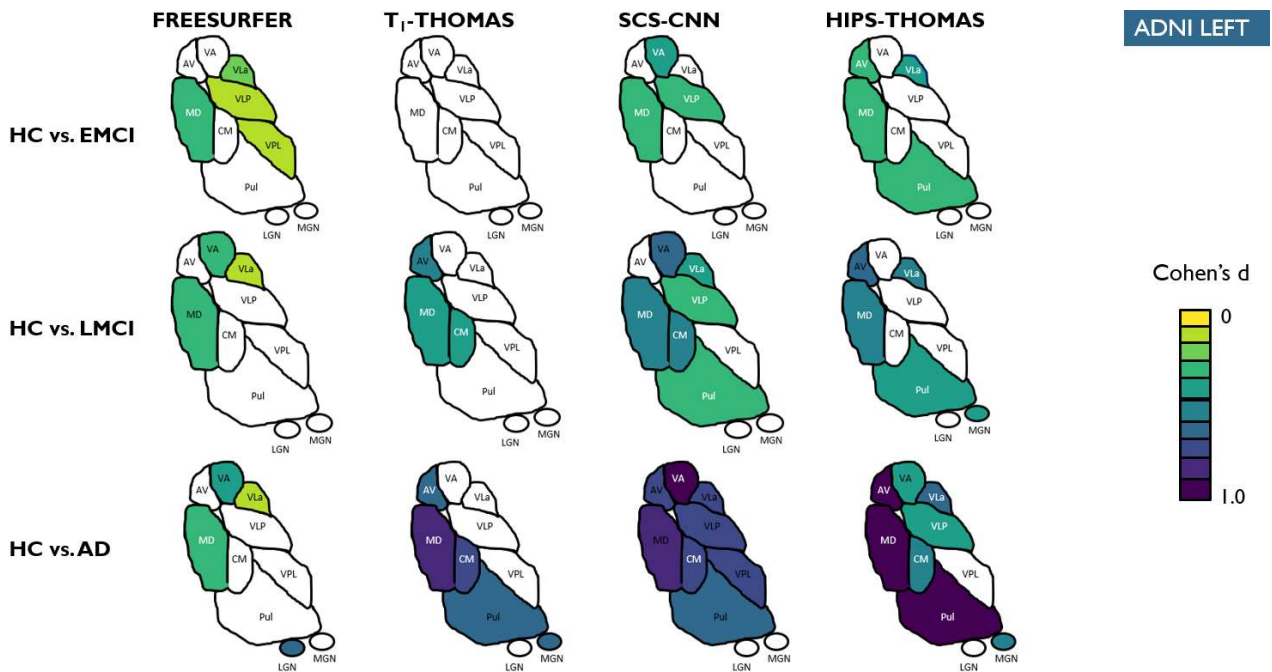

Figure 9 Left hemisphere thalamic nuclei atrophy as a function of AD stage (EMCI, LMCI, AD) for the 4 methods colorized for statistically significant nuclei using effect size (Cohen's d).

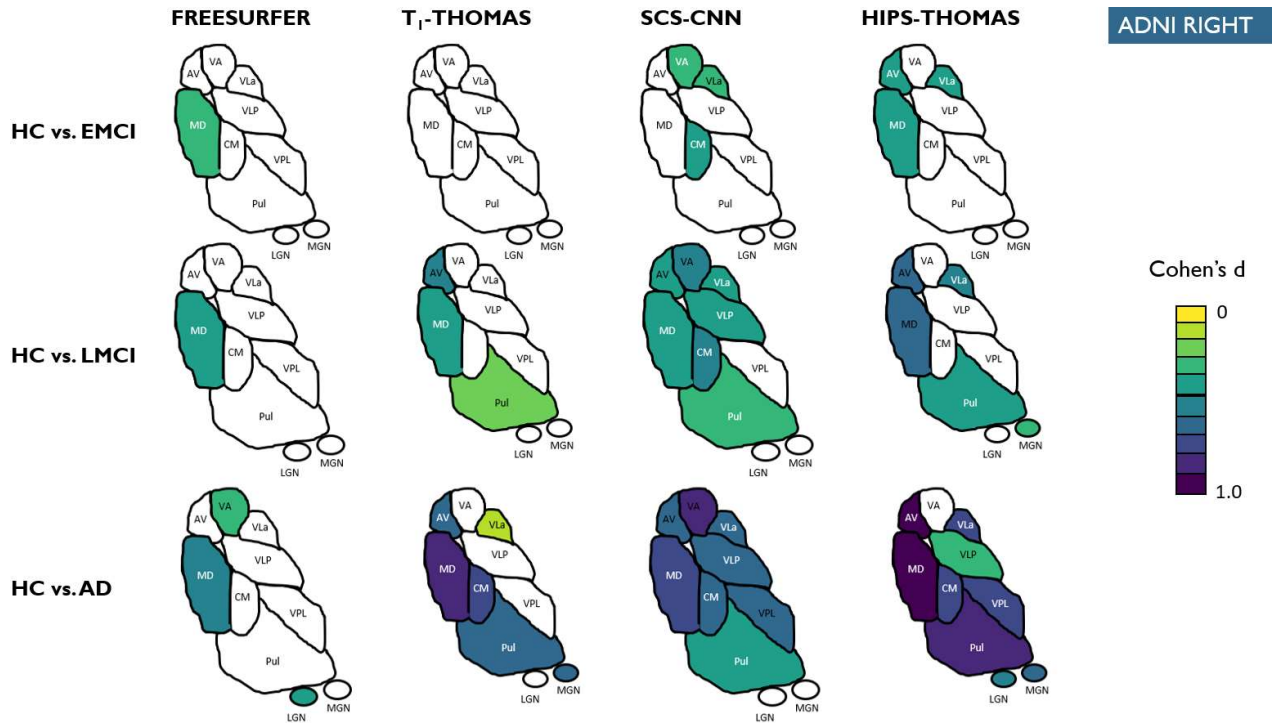

Figure 10 Right hemisphere thalamic nuclei atrophy as a function of AD stage (EMCI, LMCI, AD) for the 4 methods colorized for statistically significant nuclei using effect size (Cohen's d).

Table 4 Receiver operating characteristics analysis results. In the table below values in green highlight the greatest AUC value for discriminating healthy controls from early MCI, late MCI, and AD using either individual nuclei or whole thalamus volumes.

| Method | HC-EMCI | HC-LMCI | HC-AD |
| --- | --- | --- | --- |
| **Freesurfer** | Nuclei: 0.70<br>Whole Thal: 0.60 | Nuclei: 0.69<br>Whole Thal: 0.62 | Nuclei: 0.77<br>Whole Thal: 0.66 |
| **HIPS-THOMAS** | Nuclei: 0.74<br>Whole Thal: 0.58 | Nuclei: 0.82<br>Whole Thal: 0.63 | Nuclei: 0.93<br>Whole Thal: 0.74 |
| **SCS-CNN** | Nuclei: 0.69<br>Whole Thal: 0.62 | Nuclei: 0.76<br>Whole Thal: 0.66 | Nuclei: 0.85<br>Whole Thal: 0.76 |
| **T$_1$-THOMAS** | Nuclei: 0.73<br>Whole Thal: 0.61 | Nuclei: 0.73<br>Whole Thal: 0.65 | Nuclei: 0.81<br>Whole Thal: 0.72 |

## Discussion

To date, several thalamic nuclei segmentation methods based on diffusion MRI, resting state fMRI, and more recently, structural MRI have been reported, but comparisons across different segmentation methods are almost non-existent. While Iglehart et al. [45] qualitatively compared a method from each of those 3 classes (i.e.

diffusion, functional, and structural MRI) on a small cohort of 20 healthy subjects, this is the first work – to the best of our knowledge – that quantitatively compares state-of-the-art structural imaging-based thalamic nuclei segmentation methods, using two large cohorts to benchmark segmentation approaches using Dice and AHD coefficients, as well as AUC scores in a clinical context. For the HCP cohort, using both Dice and AHD as quantitative metrics, HIPS-THOMAS displayed the best performance (4/10 nuclei on left, 5/10 nuclei on right). These results were consistent across subject-space and MNI-space analyses. For the ADNI cohort, HIPS-THOMAS achieved the best AUC scores for discrimination between controls and all three AD disease stages – early MCI, late MCI, and AD – using individual thalamic nuclei volumes. While SCS-CNN achieved the best AUC scores using whole thalamic volumes for discrimination, these AUC values were lower than those achieved using thalamic nuclei, a result also observed for controls-AD discrimination by Iglesias et al. using the FreeSurfer Bayesian parcellation [25].

One of the main issues hampering accurate segmentation of thalamic nuclei from standard $T_1$w MRI data is the lack of intrathalamic contrast as well as the poor delineation of whole thalamic boundaries. Novel sources of contrast such as white-matter nulled contrast provided by FGATIR [46] or WMn-MPRAGE [47] significantly improve intra-thalamic contrast. They also provide a good depiction of whole thalamus boundaries, especially the ventral boundaries which are adjacent to white-matter tracts. WMn contrast [47] optimized at 7T was exploited by the original THOMAS method. The idea of synthesizing WMn-MPRAGE contrast in the absence of acquired WMn-MPRAGE data was first proposed by Datta et al. [48] who showed that WMn-MPRAGE data synthesized from the $T_1$ maps derived from the MP2RAGE acquisition improved Dice compared to direct segmentation of the MP2RAGE ratio image. Since MP2RAGE is still not commonly used at 3T and not available in public databases like HCP and ADNI, methods which directly synthesize WMn-MPRAGE-like images from $T_1$w MRI like the SCS-CNN and HIPS-THOMAS used in this work were proposed. In our analyses, both SCS-CNN and HIPS-THOMAS showed significantly improved Dice compared to FreeSurfer or $T_1$-THOMAS for the ventral nuclei suggesting the utility of improved delineation of the ventral thalamic borders enabled by the WMn contrast. The geniculate nuclei also showed significant improvements. FreeSurfer showed better Dice performance in the medial thalamus, specifically in the left mediodorsal nucleus and bilateral centromedian nucleus. One plausible reason for this difference is that because the medial thalamus shares a boundary with the third ventricle, the contrast between grey matter and CSF would be greater for traditional CSF-nulled contrast (i.e. standard MPRAGE) than WMn-MPRAGE where CSF is grey and not black. In turn, this contrast may make delineating nuclei more efficacious using standard MPRAGE as for FreeSurfer, than methods using (synthesised) WMn-MPRAGE as is the case for the SCS-CNN and HIPS-THOMAS methods. Both the WMn-synthesis-based methods (HIPS-THOMAS and SCS-CNN) achieved larger effect sizes and captured the progress of atrophy from EMCI to AD better than T1-THOMAS or Freesurfer. One major advantage HIPS-THOMAS offers over SCS-CNN (besides the larger AUC for nuclei-based discrimination) is its robustness. CNNs are very sensitive to training data and that was the case for SCS-CNN, which performed sub-optimally on Philips 3T and Siemens 7T data relative to the Siemens and GE 3T data it was trained on [33]. In contrast, the simpler polynomial based method performed more robustly on all data inputs, something which is critical when analysing large public databases which often contain a mixture of field strengths and scanner types.

Our work had several limitations. Firstly, although individual manual segmentation should be considered as the "gold" standard for creating a reference for benchmarking automated thalamic segmentation approaches, due to the impracticalities of manually labelling 640 subjects, we instead used the Krauth-Morel atlas as a surrogate "silver" standard. Secondly, due to the differences in nomenclature as well as the nuclei segmented by FreeSurfer and THOMAS plus its variants, we created a set of 10 common nuclei by merging some nuclear subdivisions such as within pulvinar and mediodorsal nuclei, to enable equivalent comparisons across methods. Thirdly, this work evaluated four of the main anatomical based segmentation methods, but did not consider all anatomical methods, or functional- or diffusion-based approaches. However, functional- and diffusion-based approaches produce segmentations that are based on statistical and diffusivity properties, respectively, and therefore are fundamentally different approaches to generating segmentations. Despite these aforementioned limitations, many published segmentation approaches make qualitative and semi-quantitative comparisons with respect to the Morel atlas. Therefore, unifying the principle anatomical segmentation methods under a single nomenclature and comparison with the Krauth-Morel atlas provides greater insight into which methods are most similar to the Morel atlas as a standard reference space.

Future work should aim to investigate how to best combine and make use of the information generated by both SCS-CNN and HIPS-THOMAS approaches, since they have complementary advantages with respect of accuracy of segmentation for different nuclei, and overall robustness. Future would also benefit from including information from other imaging modalities, as is the case for the recent improvement to the FreeSurfer based segmentation approach, which combines diffusion data with $T_1$ images [31]. Other imaging contrasts which may benefit thalamic segmentation also include magnetisation transfer which improves thalamic contrast [49] and proton density, which enhances contrast between the lateral geniculate and surrounding white matter [50]. Nevertheless, the utility of this multi-modal approach is likely to be hampered for use with publicly available secondary datasets due to a lack of availability of different imaging modalities, which are often either lacking in quality or are omitted from acquisition protocols entirely.

## Acknowledgements

This work was supported by a National Institutes of Health NIBIB Grant R01EB032674. The authors would like to thank ETH Zurich and the University of Zurich for the provision of the Morel atlas in MNI space as described in Krauth et al. (2010). Data collection and sharing for this project was partly funded by the Alzheimer's Disease Neuroimaging Initiative (ADNI) (National Institutes of Health Grant U01 AG024904) and DOD ADNI (Department of Defense award number W81XWH-12-2-0012). ADNI is funded by the National Institute on Aging, the National Institute of Biomedical Imaging and Bioengineering, and through generous contributions from the following: AbbVie, Alzheimer's Association; Alzheimer's Drug Discovery Foundation; Araclon Biotech; BioClinica, Inc.; Biogen; Bristol-Myers Squibb Company; CereSpir, Inc.; Cogstate; Eisai Inc.; Elan Pharmaceuticals, Inc.; Eli Lilly and Company; EuroImmun; F. Hoffmann-La Roche Ltd and its affiliated company Genentech, Inc.; Fujirebio; GE Healthcare; IXICO Ltd.; Janssen Alzheimer Immunotherapy Research & Development, LLC.; Johnson & Johnson Pharmaceutical Research & Development LLC.; Lumosity; Lundbeck; Merck & Co., Inc.; Meso Scale Diagnostics, LLC.; NeuroRx Research; Neurotrack Technologies; Novartis Pharmaceuticals Corporation; Pfizer Inc.; Piramal Imaging; Servier; Takeda Pharmaceutical Company; and Transition Therapeutics. The Canadian Institutes of Health Research is providing funds to support ADNI clinical sites in Canada. Private sector contributions are facilitated by the Foundation for the National Institutes of Health (www.fnih.org). The grantee organization is the Northern California Institute for Research and Education, and the study is coordinated by the Alzheimer's Therapeutic Research Institute at the University of Southern California. ADNI data are disseminated by the Laboratory for Neuro Imaging at the University of Southern California.

## Competing interests

The authors declare that they have no competing interests.

## CRediT authorship statement

Brendan Williams: Conceptualization, Methodology, Formal analysis, Resources, Data Curation, Writing - Original Draft, Writing - Review & Editing, Visualization. Dan Nguyen: Formal analysis. Julie P. Vidal: Software, Resources. Manojkumar Saranathan: Conceptualization, Methodology, Formal analysis, Resources, Data Curation, Writing - Original Draft, Writing - Review & Editing, Visualization, Supervision, Project administration, Funding acquisition.

## Data and code availability

Data were provided by the Human Connectome Project, WU-Minn Consortium (Principal Investigators: David Van Essen and Kamil Ugurbil; 1U54MH091657) funded by the 16 NIH Institutes and Centers that support the NIH Blueprint for Neuroscience Research; and by the McDonnell Center for Systems Neuroscience at Washington University. Data were also obtained from the Alzheimer's Disease Neuroimaging Initiative (ADNI) database (adni.loni.usc.edu). The ADNI was launched in 2003 as a public-private partnership, led by Principal Investigator Michael W. Weiner, MD. FreeSurfer segmentation was completed using code previously reported in Williams et al. (2022), and is available online at the University of Reading Research Data Archive (https://doi.org/10.17864/1947.000339). Code and details for running T1-THOMAS and HIPS-THOMAS using Docker are available on Github (https://github.com/thalamicseg); the Docker images are available on Docker Hub (https://hub.docker.com/u/anagrammarian). The Docker image for running SCS-CNN is available on Github (https://github.com/lunastra26/thalamic-nuclei-segmentation)

## Ethics statement

The work presented in this manuscript was conducted using open-access data from the Human Connectome Project and Alzheimer's Disease Neuroimaging Initiative.

**Supplementary methods**

**THOMAS pipeline and its variants**: The original THOMAS method that was developed and optimized for WMn-MPRAGE uses a set of 20 WMn-MPRAGE datasets ($p_1$-$p_{20}$) as priors which have been manually segmented using the Moral atlas as guide. The 20 priors are mutually registered and averaged to create a WMn template. The input image is first cropped and registered to a cropped WMn template image using ANTs nonlinear registration (R). The precomputed prior-to-template space warps ($W_{piT}$) are combined with $R^{-1}$ to warp the 20 prior labels to input space. These labels are then combined using a joint-fusion algorithm to generate a single parcellation in subject space. The WMn-MPRAGE sequence is neither part of standard clinical imaging protocols nor part of extant databases such as ADNI and OASIS. To adapt THOMAS for T1w data, one approach was to replace the cross-correlation (CC) metric with a mutual information (MI) metric in the ANTs nonlinear registration step of THOMAS and replace the joint fusion (JF) with majority voting (MV) in the label fusion step of THOMAS. We refer to this variant as T1-THOMAS. To leverage the improved intrathalamic contrast of WMn-MPRAGE, a polynomial synthesis method (box labelled HIPS) was used to first synthesize WMn-MPRAGE-like images from T1w images before applying the THOMAS algorithm. Note that the WMn-like input enables the use of the more accurate CC metric for nonlinear registration as well as the more sophisticated JF algorithm compared to MV for label fusion. We call this method HIPS-THOMAS. The original THOMAS method and the T1-THOMAS and HIPS-THOMAS variants are shown in Supplemental Figure 1 below, using green, red, and cyan colours to differentiate the three methods.

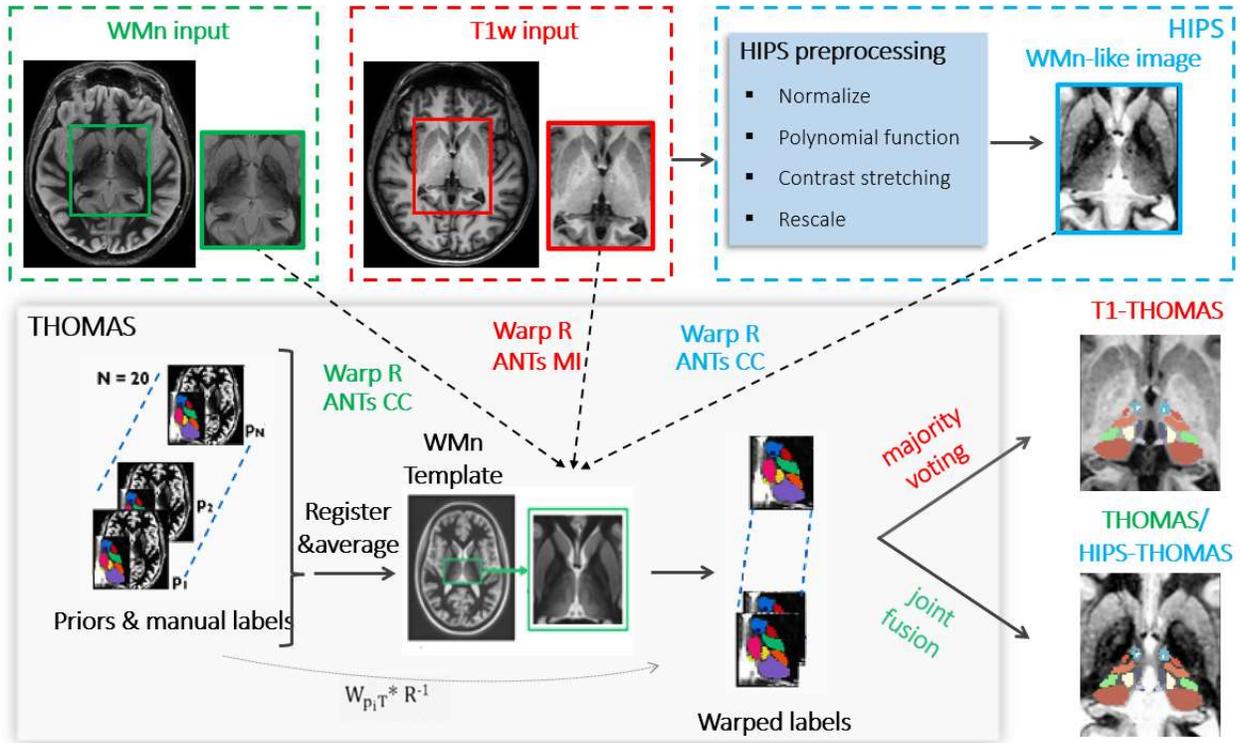

Supplementary Figure 1. Schematic of THOMAS and the two variants- T1-THOMAS and HIPS-THOMAS. T1-THOMAS (grey text) uses a mutual information metric for nonlinear registration of input to template and a majority voting algorithm to combine the labels. HIPS-THOMAS (cyan text) uses a cross-correlation metric for more accurate nonlinear registration of input to template and a joint fusion algorithm for label fusion.

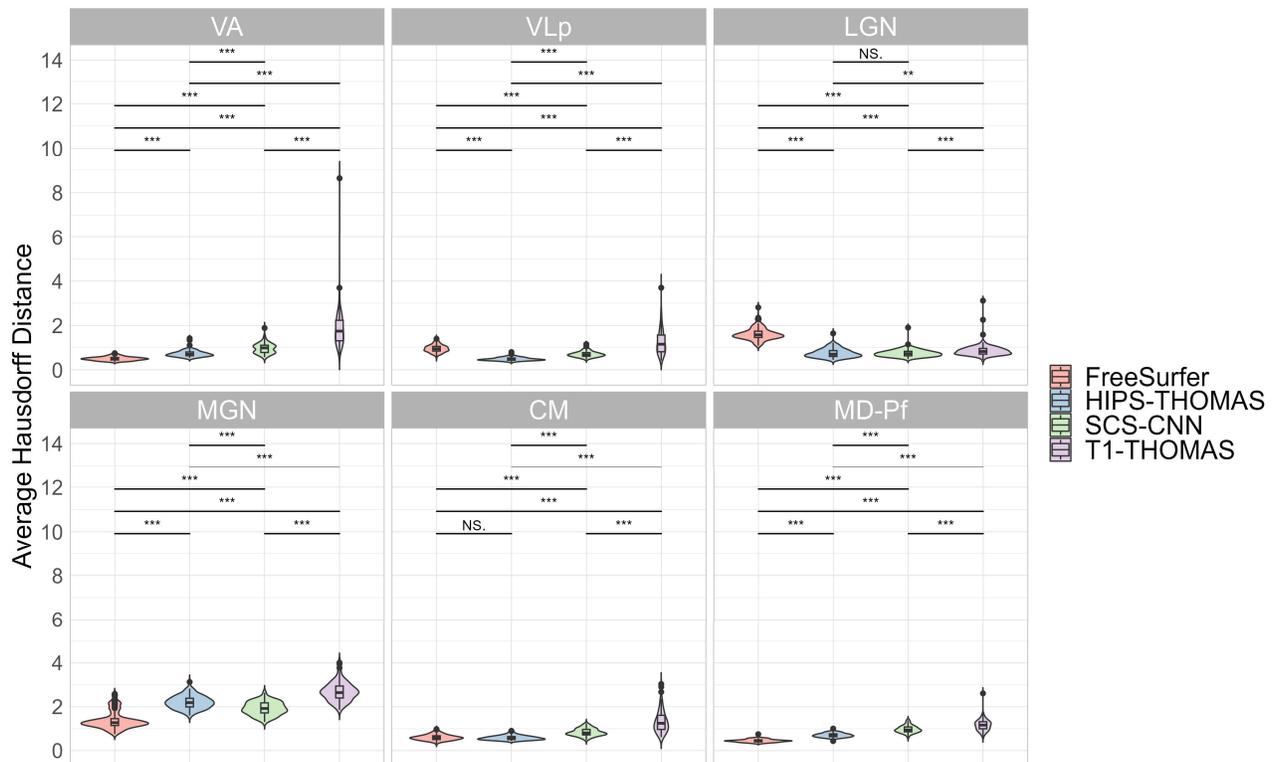

Supplementary Figure 2. Violin plots of left hemisphere nuclei with significantly different Average Hausdorff Distances for nuclei segmented from Human Connectome Project data using FreeSurfer, HIPS-THOMAS, CNN-SCS, and T1-THOMAS approaches. Posthoc t-test results (Bonferroni corrected) are presented to show pairwise difference between segmentation approaches for each nucleus (**p<0.01, ***p<0.001).

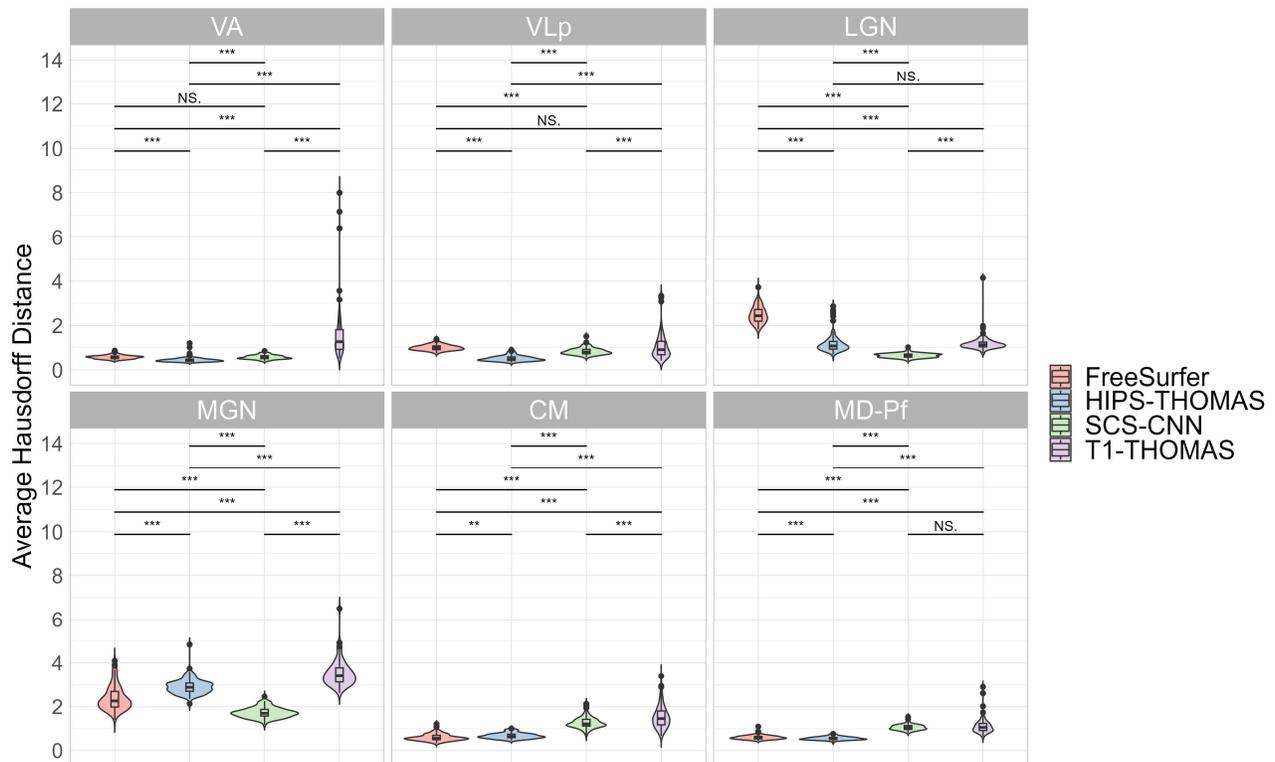

Supplementary Figure 3. Violin plots of right hemisphere nuclei with significantly different Average Hausdorff Distances for nuclei segmented from Human Connectome Project data using FreeSurfer, HIPS-THOMAS, CNN-SCS, and T1-THOMAS approaches. Posthoc t-test results (Bonferroni corrected) are presented to show pairwise difference between segmentation approaches for each nucleus (**p<0.01, ***p<0.001).

Supplementary table 1. Two-way ANOVA results for HCP dataset analysis in subject space for each nucleus. Significant main effects of segmentation approach (dataset) and hemisphere (side), and interactions were found for all nuclei except for VPL, which did not show a main effect of side.

| Effect | DFn | DFd | F | p | p<.05 | ges | segmentation |
|---|---|---|---|---|---|---|---|
| Dataset | 2.56 | 251.05 | 138.753 | 3.06E-48 | * | 0.333 | AV |
| side | 1 | 98 | 156.403 | 5.10E-22 | * | 0.196 | AV |
| Dataset:side | 1.9 | 185.8 | 110.71 | 2.50E-31 | * | 0.199 | AV |
| Dataset | 1.35 | 132.2 | 596.277 | 1.24E-57 | * | 0.719 | VA |
| side | 1 | 98 | 286.446 | 7.60E-31 | * | 0.213 | VA |
| Dataset:side | 1.83 | 179.69 | 100.781 | 1.75E-28 | * | 0.126 | VA |
| Dataset | 2.11 | 206.35 | 85.01 | 1.30E-28 | * | 0.31 | VLa |
| side | 1 | 98 | 79.947 | 2.40E-14 | * | 0.059 | VLa |
| Dataset:side | 1.97 | 192.81 | 166.994 | 2.08E-42 | * | 0.251 | VLa |
| Dataset | 1.56 | 153.36 | 184.886 | 1.79E-36 | * | 0.498 | VLp |
| side | 1 | 98 | 18.759 | 3.59E-05 | * | 0.016 | VLp |
| Dataset:side | 1.47 | 143.99 | 55.408 | 3.26E-15 | * | 0.08 | VLp |
| Dataset | 1.72 | 168.96 | 75.557 | 5.91E-22 | * | 0.275 | VPL |
| side | 1 | 98 | 0.692 | 0.407 | | 0.000709 | VPL |
| Dataset:side | 1.79 | 175.34 | 56.634 | 2.78E-18 | * | 0.1 | VPL |
| Dataset | 2.74 | 268.54 | 407.59 | 1.34E-95 | * | 0.593 | LGN |
| side | 1 | 98 | 357.482 | 1.83E-34 | * | 0.267 | LGN |
| Dataset:side | 2.57 | 252.24 | 92.152 | 1.79E-36 | * | 0.202 | LGN |
| Dataset | 1.25 | 122.32 | 188.742 | 5.26E-30 | * | 0.515 | MGN |
| side | 1 | 98 | 698.738 | 2.18E-46 | * | 0.35 | MGN |
| Dataset:side | 1.47 | 143.66 | 126.936 | 3.58E-27 | * | 0.2 | MGN |
| Dataset | 2.07 | 202.89 | 490.928 | 1.16E-79 | * | 0.531 | Pul |
| side | 1 | 98 | 268.92 | 7.54E-30 | * | 0.23 | Pul |
| Dataset:side | 1.56 | 152.6 | 35.716 | 2.16E-11 | * | 0.048 | Pul |
| Dataset | 1.65 | 161.51 | 523.075 | 7.27E-66 | * | 0.74 | CM |
| side | 1 | 98 | 277.192 | 2.52E-30 | * | 0.127 | CM |
| Dataset:side | 2.09 | 205.2 | 80.828 | 1.95E-27 | * | 0.101 | CM |
| Dataset | 1.62 | 159.22 | 641.411 | 5.30E-71 | * | 0.758 | MD-Pf |
| side | 1 | 98 | 37.576 | 1.85E-08 | * | 0.036 | MD-Pf |
| Dataset:side | 1.58 | 155.05 | 147.22 | 5.01E-32 | * | 0.202 | MD-Pf |

Supplementary table 2 rankings for AHD metrics for nuclei with main effects of segmentation approach for HCP data in subject space

| Nucleus | Left | | | | Right | | | |
| --- | --- | --- | --- | --- | --- | --- | --- | --- |
| | FreeSurfer | HIPS-THOMAS | SCS-CNN | T1-THOMAS | FreeSurfer | HIPS-THOMAS | SCS-CNN | T1-THOMAS |
| VA | 1 | 2 | 3 | 4 | 2 | 1 | 2 | 4 |
| VLp | 3 | 1 | 2 | 4 | 3 | 1 | 2 | 3 |
| LGN | 4 | 1 | 1 | 3 | 4 | 2 | 1 | 2 |
| MGN | 1 | 3 | 2 | 4 | 2 | 3 | 1 | 4 |
| CM | 1 | 1 | 3 | 4 | 1 | 2 | 3 | 4 |
| MD-Pf | 1 | 2 | 3 | 4 | 2 | 1 | 3 | 3 |
| Mean | 1.833 | 1.667 | 2.333 | 3.833 | 2.333 | 1.667 | 2 | 3.333 |
| SD | 1.213 | 0.745 | 0.745 | 0.373 | 0.943 | 0.745 | 0.816 | 0.745 |